\documentclass[a4paper,fleqn,usenatbib,useAMS]{mnras}
\usepackage{graphics}
\usepackage{graphicx}
\usepackage[pagewise]{lineno}
\usepackage{amsmath}    
\usepackage{amssymb}    
\usepackage{multicol}        
\usepackage{bm}     
\usepackage{pdflscape}  
\usepackage{xcolor}
\usepackage{ae,aecompl}
\usepackage{subfigure}
\usepackage{scalerel}
\usepackage[english]{babel}
\usepackage{times}
\usepackage{physics}
\usepackage{etoolbox}
\usepackage{multirow, booktabs}
\usepackage{siunitx}
\usepackage{paralist}
\usepackage{makecell}
\usepackage{tabularx}
\usepackage[inline]{enumitem}
\newlist{mycompactenum}{enumerate}{1}
\setlist[mycompactenum,1]{nosep,label=\arabic*.}
\usepackage{paralist}

\newcommand{\wfirst}{\textit{Roman}}

\title[The \wfirst~sensitivity to planets in the HZs]{Sensitivity to habitable planets in the \wfirst~microlensing survey}
\author[Sajadian]{Sedighe Sajadian$~^{1,2}$\thanks{E-mail: s.sajadian@iut.ac.ir}\\
	$^{1}$Department~of~Physics,~Isfahan~University~of~Technology,~Isfahan~84156-83111,~Iran\\
	$^{2}$Department~of~Physics,~Chungbuk~National~University,~Cheongju~28644,~Republic~of~Korea}

\begin{document}

\label{firstpage}
\pagerange{\pageref{firstpage}--\pageref{lastpage}}
\maketitle
\begin{abstract}	
We study the \wfirst~sensitivity to exoplanets in the Habitable Zone (HZ). The \wfirst~efficiency for detecting habitable planets is maximized for three classes of planetary microlensing events with close caustic topologies. (a) The events with the lens distances of $D_{\rm l} \gtrsim 7$ kpc, the host lens masses of $M_{\rm h}\gtrsim 0.6M_{\odot}$. By assuming Jupiter-mass planets in the HZs, these events have $q\lesssim 0.001$ and $d\gtrsim 0.17$ ($q$ is their mass ratio and $d$ is the projected planet-host distance on the sky plane normalized to the Einstein radius). The events with primary lenses, $M_{\rm h} \lesssim 0.1 M_{\odot}$, while their lens systems are either (b) close to the observer with $D_{\rm l}\lesssim 1$ kpc or (c) close to the Galactic bulge, $D_{\rm l}\gtrsim 7$ kpc. For Jupiter-mass planets in the HZs of the primary lenses, the events in these two classes have $q\gtrsim 0.01$, $d\lesssim 0.04$. The events in the class (a) make larger caustics. By simulating planetary microlensing events detectable by \wfirst,~we conclude that the \wfirst~efficiencies for detecting Earth- and Jupiter-mass planets in the Optimistic HZs (OHZs, which is the region between $[0.5,~2]$ AU around a Sun-like star) are $0.01\%$ and $5\%$, respectively. If we assume that one exoplanet orbits each microlens in microlensing events detectable by \wfirst~( i.e., $\sim 27000$ ),~this telescope has the potential to detects $35$ exoplanets with the projected planet-host distances in the OHZs with only one having a mass $\lesssim 10M_{\oplus}$. According to the simulation, $27$ of these exoplanets are actually in the OHZs.
\end{abstract}

\begin{keywords}
Gravitational lensing: micro; planets and satellites: detection; planets and satellites: terrestrial planets; methods: numerical. 
\end{keywords}

\section{Introduction}

One of the primary motivations for searching for extrasolar planets is to look for extraterrestrial life outside of our solar system and to resolve the Fermi paradox \citep[see, e.g., ][]{FermiPara1998, Fermiparad, Fermiparad2015, 2020Fermi}. According to this paradox, the likelihood of extraterrestrial civilizations elsewhere in our galaxy is high, despite the fact that no evidence of them has been found. Indeed, there are many planetary systems in our galaxy similar to our solar system. Some of them must be suitable for developing intelligent beings and we should have received signals from them up to now \citep[e.g., see,][]{Huang1959, 2013Kmar}.  

The existence of a liquid bio-solvent is one of the necessary factors for life. Liquid water is the best candidate that exits in a wide temperature range, i.e., $[273,~373]$ K \citep[][]{Huang1959, Kasting1993, HabitPlanet2019, Forget2013}. The range of orbits around each star that can support liquid water over the surface of its orbiting planets is the so-called 'Habitable Zone' (HZ). However, the existence of liquid water additionally needs sufficient atmospheric pressure over the planet's surface. 

\noindent The equilibrium temperature over the planet's surface, $T_{\rm p}$, is determined by the Bond albedo, $A_{\rm B}$, which is the fraction of the incident radiation to the planet's surface that is scattered back into space, and the re-radiation factor, $f$, and is given by \citep[see, e.g., ][]{2014Peters, 2017Catling}:  
\begin{eqnarray}
T_{\rm p}=T_{\ast}\sqrt{\frac{R_{\ast}}{s}}\Big[f (1-A_{\rm B})\Big]^{1/4}
\end{eqnarray}
where, $T_{\ast}$ and $R_{\ast}$ are the effective temperature and the radius of the host star, respectively, $s$ is the planet-host distance. Here, the planet is considered as a black body being heated only by its host star. For rocky planets, the Bond albedo and re-radiation factor are $A_{\rm B}\simeq 0.2$ and $f\simeq 0.35$ which result the Earth temperature equals to $T_{\oplus}\simeq 289$ K.

Two kinds of HZs are defined (narrower and wider zones): Conservative (CHZ) and Optimistic (OHZ). In our solar system, the CHZ and OHZ represent the circumstellar zones with boundaries $[0.95,~1.67]$ AU and $[0.5,~2]$ AU (including Venus and Mars orbits), respectively \citep{Kopparapu2013, Kopparapu2014, Chandler2016}. In the CHZ and OHZ of the Sun, the surface temperature of orbiting planets can be between $[228,~294]$ K and $[203,~405]$ K, respectively. In order to determine the outer and inner boundaries of the OHZ, a thick atmosphere with a strong greenhouse effect and a thin atmosphere with a weak greenhouse effect are assumed, respectively. Therefore, the first step to probing extraterrestrial life is searching for extrasolar planets whose orbits stay in the HZs, such as the Earth in our solar system. 

Up to now, more than $4017$ extrasolar planets in $3016$ planetary systems have been detected, according to the Habitable Zone Gallery\footnote{\url{http://www.hzgallery.org}} \citep{Kane2012b}. Among them, there are $141$ exoplanets with orbits entirely within the OHZs, the so-called habitable planets. The majority of these exoplanets have been discovered via radial velocity measurements or transit methods \citep[e.g., ][]{Borucki2012, Gilbert2020}. One of these habitable planets has been discovered through microlensing observations which were done by the Microlensing Observations in Astrophysics (MOA) microlensing group \citep{moa2001, Sumi2003, HZPmicrolensing}.

The highest sensitivity to exoplanet detection through microlensing observations occurs on planets beyond the snow line. These planetary systems frequently make resonant or close-to-resonant caustic topologies \citep[e.g., ][]{1998GriestSafizadeh, DiStefano1999, Park2006,Dominik2008,gaudi2012, 2016sajadian,2021Yee}. For that reason, only one habitable exoplanet, in the microlensing event MOA-2011-BLG-293Lb, has been discovered through microlensing observations up to now \citep{HZPmicrolensing, Yee2012}. However, the recent analysis of this event, based on high-angular resolution follow-up imaging observations, has shown that the excess flux at the source position in this event likely does not come from the host microlens object \citep{2020Koshimoto}. Their results suggest a lower mass for the host star.

Habitable planetary systems most often form binary-lens systems with close caustic topologies characterized by three small caustic curves. The probability of passing through these small caustics or covering such short caustic-crossing features is low. Nonetheless, the new generation of telescopes and microlensing surveys, such as \textit{The Nancy Grace Roman Space Telescope} (\wfirst)~survey with the improved cadence (i.e., $\simeq 15$ min) and better resolutions potentially may discover more habitable planets through microlensing observations. However, one of the main goals of the \wfirst~Galactic Exoplanet Survey is to detect a large sample of free-floating or bound, cool and low-mass exoplanets and measure the mass functions of these planets, which differs somewhat from searching exclusively for exoplanets in the HZs. We emphasize that the microlensing method can detect planetary systems at farther distances with respect to the observer (even in the Galactic bulge) than those detectable through transit or radial velocity measurements. 

In this paper, we first revisit the subject of probing habitable planets through microlensing observations. Then, we will do a Monte-Carlo simulation of planetary microlensing events to evaluate the \wfirst~efficiency for detecting habitable planets and probe the special microlensing configurations with the maximum \wfirst~sensitivity to habitable planets. Finally, we calculate how many habitable planets could be discovered during the mission.

The paper is organized as follows. In section \ref{two}, we explain some points for detecting habitable planets through microlensing observations. In section \ref{three}, we study the properties and statistics of the realizable microlensing events owing to habitable planetary systems by \wfirst. In the two next sections, \ref{four} and \ref{five}, we discuss on and summarize the results. 
\begin{figure}
\centering
\includegraphics[width=0.49\textwidth]{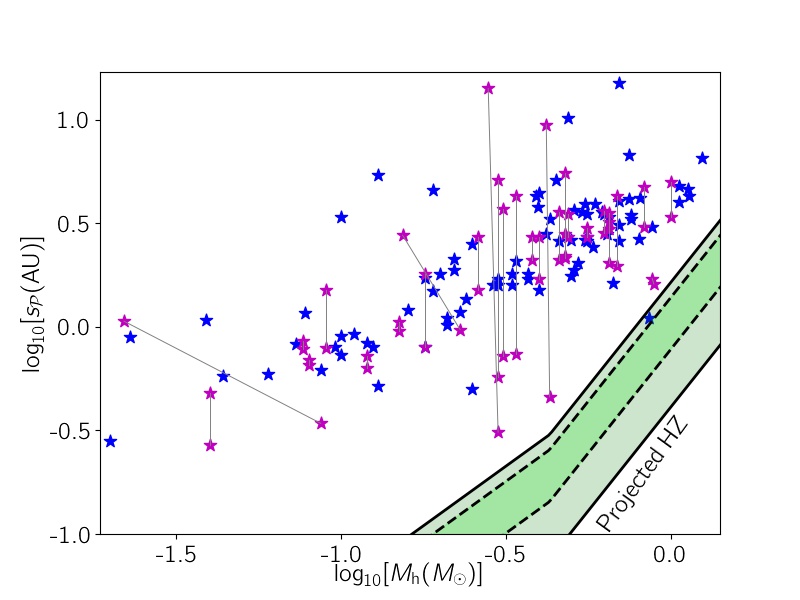}
\caption{The scatter plot of confirmed exoplanets through microlensing observations up to now. The data was taken from the NASA Exoplanet Archive. The blue stars represent the events with non-degenerate models. For the other events, shown with magenta stars, there are two (or more) degenerate models which are connected with gray lines. The solid and dashed black lines represent the boundaries of the projected OHZs and CHZs, respectively. The taken projection factor is $f_{\mathcal{P}}=0.79$, as given by Eq. \ref{projected}.}\label{fig1}
\end{figure}

\section{Detecting habitable planetary systems by microlensing} \label{two}
The subject of "microlensing abilities for detecting habitable planets" has been studied in several references. \citet{DiStefano1999} has noticed that habitable Earth-mass planets around Sun-like stars make most often resonant caustic configurations with the highest sensitivity. In this reference, the author has studied the advantages and disadvantages of the blending effect of these bright host stars. The sensitivity to habitable Earth-mass planets through microlensing observations has also been studied by \citet{Park2006} and they concluded that roughly $3\%$ of Earth-mass planets detectable in microlensing observations are inside the HZs. We note that they have used an estimation of the caustic size as the criterion for detecting planetary signatures. This choice caused an overestimation for detectable habitable planets. In this section, I theoretically revisit the mentioned subject. 
\begin{figure*}
\centering
\includegraphics[width=0.49\textwidth]{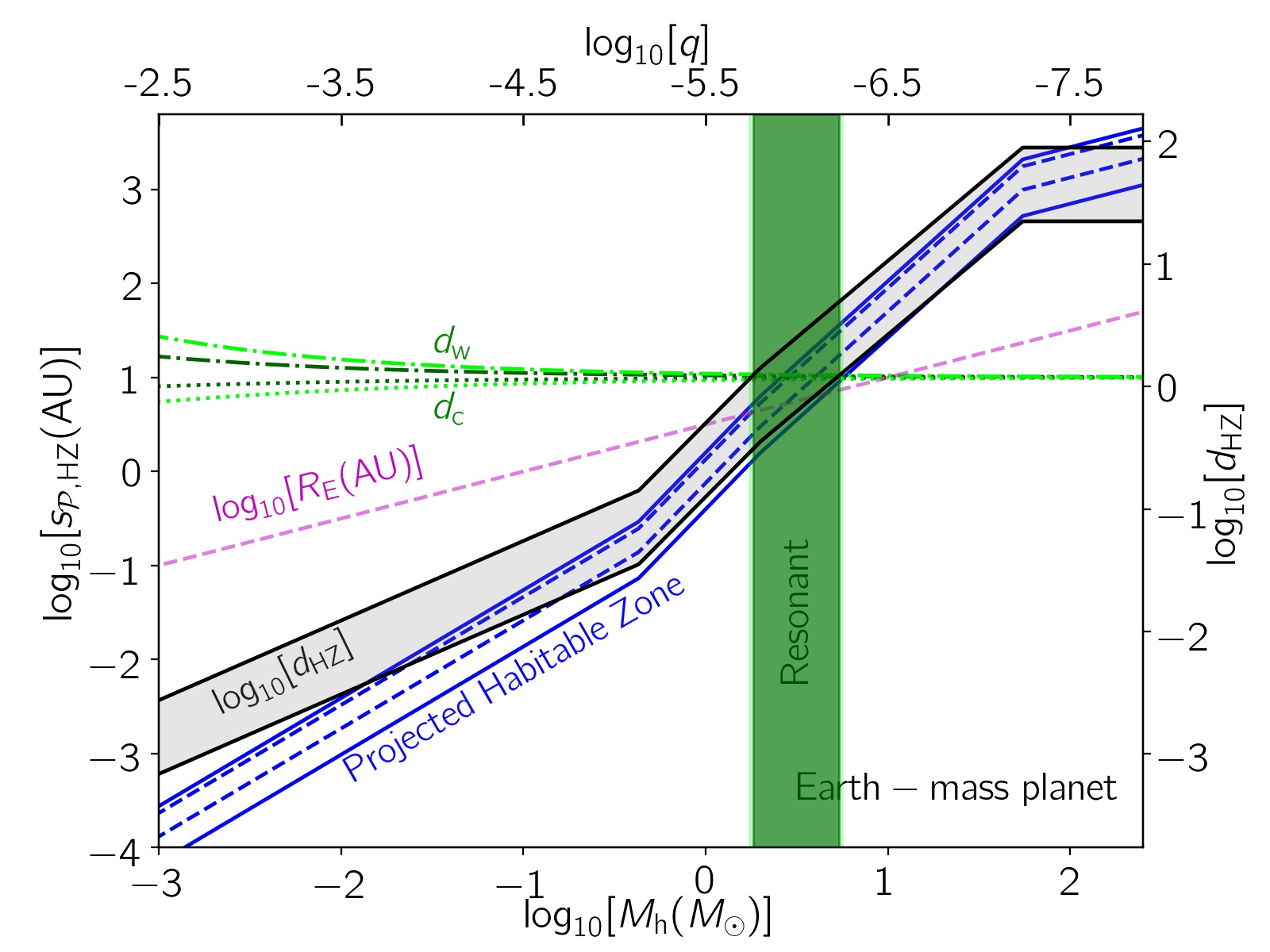}
\includegraphics[width=0.49\textwidth]{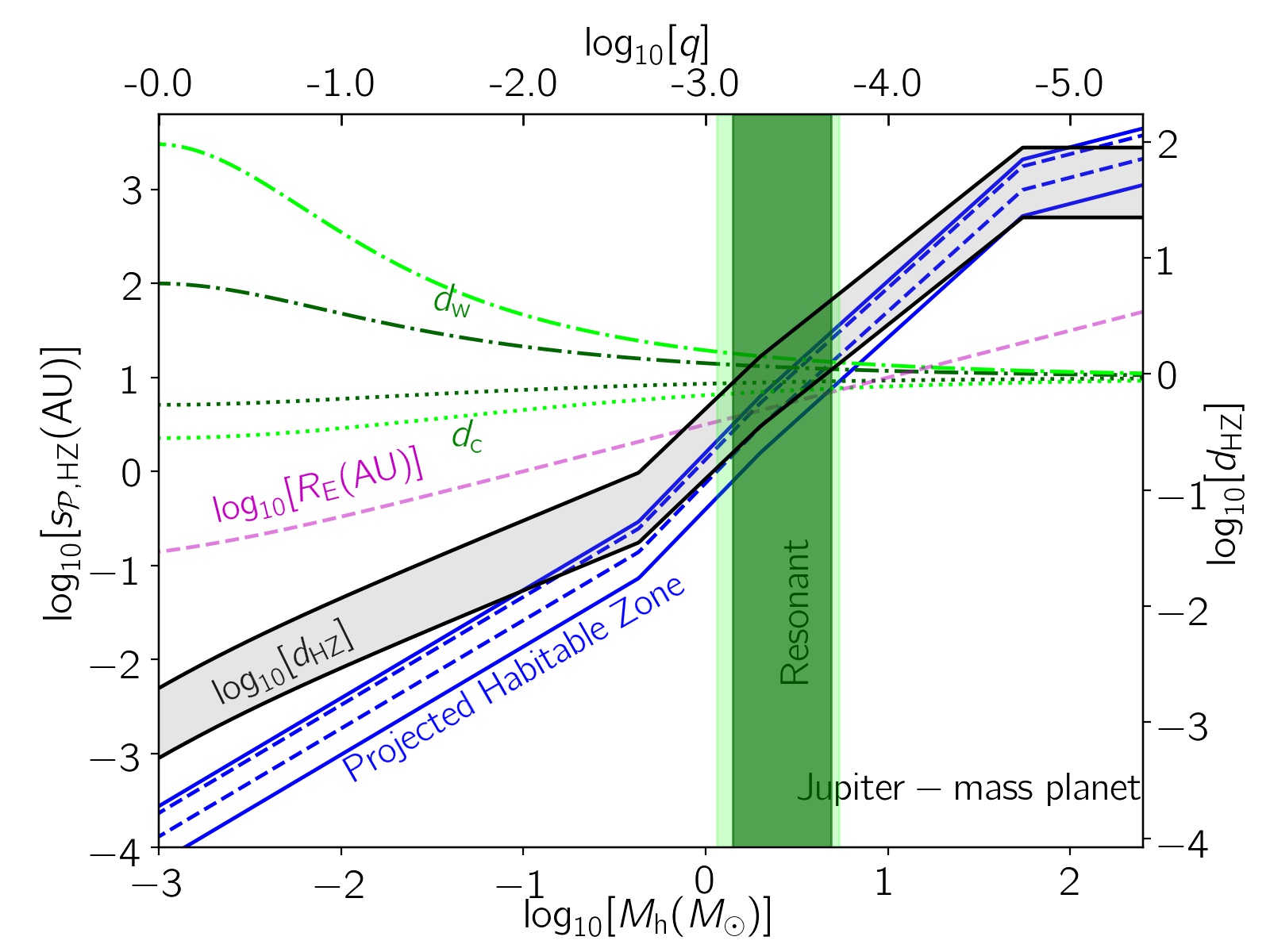}
\caption{In these plots, the OHZ and CHZ boundaries projected on the sky plane, $s_{\mathcal{P},~\rm{HZ}}(\rm{AU})$, versus the host mass are represented with solid and dashed blue lines, respectively. By assuming Earth-mass (left panel) and Jupiter-mass (right panel) planets orbiting each host star as microlens systems, their typical Einstein radii (with $D_{\rm l}=6.5$ kpc, $D_{\rm s}=8$ kpc) are shown with the magenta dashed line. The edges of projected OHZs normalized to the Einstein radius, $d_{\rm{HZ}}$, (specified on the second $y$-axis) are plotted with solid black lines. Accordingly, the habitable planetary systems inside the green-colored (lime-colored) parts make resonant (close-to-resonant) caustic topologies. To better providing, the bifurcation values between close, resonant and wide topologies, $d_{\rm c}$ and $d_{\rm w}$ (given by Eq. \ref{bifurcation}), are displayed with dotted and dot-dashed dark-green lines. Their corresponding values due to close-to-resonant topologies \citep{2021Yee} are shown with lime color.}\label{HZppp}
\end{figure*}

Up to now, more than $100$ exoplanets have been confirmed by surveys and follow-up microlensing observations, as they have been archived in the Extrasolar Planets Encyclopedia \footnote{\url{http://exoplanet.eu/}}. We take the parameters of these planets from the NASA Exoplanet Archive \footnote{\url{https://exoplanetarchive.ipac.caltech.edu/}}. In Figure \ref{fig1} the distribution of these exoplanets is shown. For some of these microlensing events, two or more best-fitted and degenerate models have been reported, which are shown by magenta stars in this plot. For each microlensing event, its degenerate models are connected with a gray line. Other discovered planetary microlensing events with non-degenerate confirmed models are displayed with blue stars. In this plot, the axes are the planet-host distance projected onto the sky plane $s_{\mathcal{P}} (\rm{AU})$ (the index $\mathcal{P}$ refers to the projection on the sky plane) and the mass of the host star $M_{\rm h}$.  

\noindent As explained in the previous section, the inner and outer radii of the CHZ and OHZ around a Sun-like star are $s_{\rm{HZ,~i,~\odot}}=0.95,~0.5$ AU and $s_{\rm{HZ,~o,~\odot}}=1.67,~2.0$ AU, respectively \citep[e.g., ][]{Kopparapu2013}. Thereupon, for a host star with the luminosity $L_{\rm h} (L_{\odot})$  the inner and outer edges of its habitable zone are given by:
\begin{eqnarray}
s_{\rm{HZ},~i}(\rm{AU}) &=& s_{\rm{HZ},~i,~\odot} \sqrt{L_{\rm h} (L_{\odot} ) }, \nonumber \\
s_{\rm{HZ},~o}(\rm{AU})&=& s_{\rm{HZ},~o,~\odot} \sqrt{L_{\rm h} (L_{\odot} ) }.
\end{eqnarray}

\noindent In Figure \ref{fig1}, we show the projected CHZ and OHZ edges with dashed and solid black lines. For the projection on the sky plane, we multiply the edges by the projection factor,  i.e., $s_{\mathcal{P},~\rm{HZ}}= f_{\mathcal{P}}~s_{\rm{HZ}}$, and $f_{\mathcal{P}}= 0.79$, as given by Eq. \ref{projected}.

\noindent In this figure, we use the known mass-luminosity relation for main-sequence stars \citep[e.g., ][]{Kuiper1938} to determine the HZ edges for a given host mass. This relation is \citep{booksalaris,bookDuric}:  
\begin{eqnarray}
L_{\rm h}(L_{\odot})= \nonumber~~~~~~~~~~~~~~~~~~~~~~~~~~~~~~~~~~~~~~~~~~~~~~~~~~~~~~~~~~~~~~~~~ \\
\begin{cases}
0.23 \left[M_{\rm h}(M_{\odot})\right]^{2.3} &  M_{\rm h}(M_{\odot})\leq 0.43,  \\ 
\left[M_{\rm h}(M_{\odot})\right]^{4}  &  0.43<M_{\rm h}(M_{\odot})\leq 2,\\ 
1.4 \left[M_{\rm h}(M_{\odot})\right]^{3.5}  &  2<M_{\rm h}(M_{\odot})\leq 55,\\
32000 M_{\rm h}(M_{\odot}) &  M_{\rm h}(M_{\odot})>55 .
\end{cases}\label{massL}
\end{eqnarray}

\noindent The only habitable planet discovered in microlensing observations, i.e., MOA-2011-BLG-293Lb, is located over the green area in the plot. This planetary system contains a super-Jupiter orbiting an M dwarf with the mass $M_{\rm h}=0.86~M_{\odot}$ at the distance $7.7$ kpc from us, close to the Galactic bulge \citep{HZPmicrolensing}. The reason for such a small number of habitable planets detected by gravitational microlensing is that its sensitivity is maximized for planets beyond the snow lines (out of the HZs). These planetary systems make intermediate (resonant) or close-to-intermediate (close-to-resonant) caustic curves \citep{2021Yee}.

In Figure \ref{HZppp}, we show the OHZ and CHZ boundaries projected on the sky plane, $s_{\mathcal{P},~\rm{HZ}} (\rm{AU})$, versus the mass of the host star with solid and dashed blue lines, respectively. By assuming Earth-mass (left panel) and Jupiter-mass (right panel) planets orbiting host stars as microlens systems, we calculate typical values of their Einstein radius (with the lens and source distances from the observer $D_{\rm l}=6.5$ kpc, $D_{\rm s}=8$ kpc, respectively), as shown with the magenta dashed lines. The edges of projected OHZs normalized to the Einstein radius, $d_{\rm{HZ}}=s_{\mathcal{P},~\rm{HZ}}/R_{\rm E}$, (specified on the second $y$-axis) are plotted with solid black lines.
		
\noindent These habitable planetary systems make close, intermediate or wide caustic topologies. We determine these topologies using two bifurcation values of the projected planet-host distance normalized to the Einstein radius, (i) $d_{\rm c}$, between close and intermediate caustics, and (ii) $d_{\rm w}$ which determines the transition between intermediate and wide caustic topologies. These two bifurcation values depend only on the planet-host mass ratio $q$ as follows \citep{1993Erdlschneider, 1999Dominikbin, 2008Cassan, 2018Tsapras}:
\begin{eqnarray}
d_{\rm c}^{8}&=& \frac{\big(1+q)^{2}}{27 q}\Big(1-d_{\rm c}^4\Big)^{3}, \nonumber\\
d_{\rm w}^{2}&=& \frac{\Big(1+ q^{1/3}\Big)^{3}}{1+q}.  \label{bifurcation}
\end{eqnarray}

\noindent In Figure \ref{HZppp}, we plot the bifurcation values, $d_{\rm c}$ and $d_{\rm w}$, with dotted and dot-dashed dark-green lines. Recently, \citet{2021Yee} have shown that majority of discovered planetary microlensing events are found inside a region between $d_{\rm c}^{3}$ and $d_{\rm w}^{1.8}$, the so-called close-to-resonant (or near-resonant) zone. These two new edges (due to near-resonant region) are shown with lime color. Accordingly, habitable planetary systems located in the green-colored (lime-colored) parts of these plots make resonant (close-to-resonant) caustic topologies. For these planetary systems in resonant (green-colored) part, some values of $d_{\rm HZ}$ (between two solid black lines) are inside the range $[d_{\rm c},~d_{\rm w}]$.
	
\noindent In this figure we use a typical value for the lens distance $D_{\rm l}= 6.5$ kpc, but the lens object can be very close to the observe or very close to the source star. Hence, we change the lens distance in the range $D_{\rm l} \in [0.08,~7.92]$ kpc and study the properties of detectable habitable planetary systems in different distances from the observer through microlensing observations. Accordingly, we list some key points in the following.
\begin{itemize}

\item Habitable planets around host stars with $M_{\rm h} \lesssim 0.5 M_{\odot}$, generate close caustic topologies, for all values of the lens distance in the range $D_{\rm l} \in [0.08,~7.92]$ kpc. However, the size of these planetary caustics depends on the mass ratio $q$. For Earth and Jupiter-mass planets around main-sequence stars, the resulting mass ratios are $\log_{10} [q]\lesssim -5.0,~-2.5$, respectively (as given on the second $x$ axes of Figure \ref{HZppp}). We note that Jupiter-mass planets can have habitable moons orbiting them.

\item When $s_{\mathcal{P},~\rm{HZ}} \simeq R_{\rm E}$, the planetary system makes resonant caustic curves, as specified with green-colored zone for the lens distance $D_{\rm l}=6.5$ kpc in Figure \ref{HZppp}. Generally, these configurations happen for the host stars with the mass of $M_{\rm h} \in [0.64,~5.62] M_{\odot}$ and $M_{\rm h} \in [0.61,~6.03] M_{\odot}$ for Earth- and Jupiter-mass planets, respectively (by assuming $D_{\rm l}\in [0.08,~7.92]$ kpc). We note that these intermediate caustic topologies happen for some special values of $s_{\mathcal{P},~\rm{HZ}}$ and $D_{\rm l}$, but not for all values. For instance, a Jupiter-mass planet in the HZ of the host star with $M_{\rm h} =0.6~M_{\odot}$ should be at the distances $D_{\rm l}\simeq 7.92$ kpc to produce intermediate caustic curves.  
 
\item When the lens system is very close to or very far from the observer, i.e., $x<0.1$ or $x>0.9$, where $x=D_{\rm l}/D_{\rm s}$, and for host stars with $M_{\rm h} \gtrsim 0.6 M_{\odot}$ (orbiting by Earth or Jupiter-mass planets in the HZs) the Einstein radius decreases and the intermediate caustic topologies form. However, for very close lens systems with $x<0.1$, the blending effect of the bright host star increases. Hence, very far lens systems with $x>0.9$ are more detectable and they occur more frequently. We note that the habitable planetary system discovered by the MOA microlensing observations, i.e., MOA-2011-BLG-293Lb, is at a distance $D_{\rm l}=7.7$ kpc from us with $x \simeq 0.96$.    

\item Although the more massive host stars have larger Einstein radii, the HZs around them are at larger distances as well. According to the mass-luminosity relation, given by Equation \ref{massL}, the projected planet-host distance in the HZs normalized to the Einstein radius, $d_{\rm{HZ}}$, is proportional to:
\begin{eqnarray}
d_{\rm{HZ}}=\frac{s_{\mathcal{P},~\rm{HZ}}}{R_{\rm E}}\propto
\begin{cases} 
\left[M_{\rm h}(M_{\odot})\right]^{0.65},& M_{\rm h}(M_{\odot})\leq 0.43,\\
\left[M_{\rm h}(M_{\odot})\right]^{1.5}, & 0.43<M_{\rm h}(M_{\odot})\leq 2. \\
\end{cases} 
\end{eqnarray}
Hence, more massive host stars have larger normalized lens distances and, as a result, bigger caustic curves.  

\item The size of the Einstein radius for the events toward the Galactic bulge is $R_{\rm{E}}\simeq 2.2 $ AU, (as specified over the $y$-axis of the plots) and primarily larger than the HZ location. One way to decrease the Einstein radius is microlensing observations toward the Galactic disc. In the Galactic disc, the source distance is on average less than $8$ kpc (the Galactic center distance), resulting in a smaller Einstein radius \citep[e.g., ][]{Moniez2017}. We expect higher efficiency for detecting planets within the HZs towards the Galactic disc than that for the Galactic bulge. In addition, towards the Galactic disc, the lens-source relative velocity is small and the Einstein crossing times are longer \citep[e.g., see Fig. (4) of ][]{Sajadian2019}.  
\end{itemize}

In order to evaluate the \wfirst~ability to discern habitable planets in its microlensing observations toward the Galactic bulge, we perform a Monte-Carlo simulation by taking into account all relevant factors, e.g., the blending effect of the lens, observing cadence, seasonal gap, photometric accuracy, etc. Based on this simulation, we access the likelihood of discerning habitable planets using the next generation of microlensing observation by the \wfirst~telescope and study the properties of these detectable events. This simulation and its results are explained in the following section. 

\begin{figure*}
\centering
\includegraphics[width=0.49\textwidth]{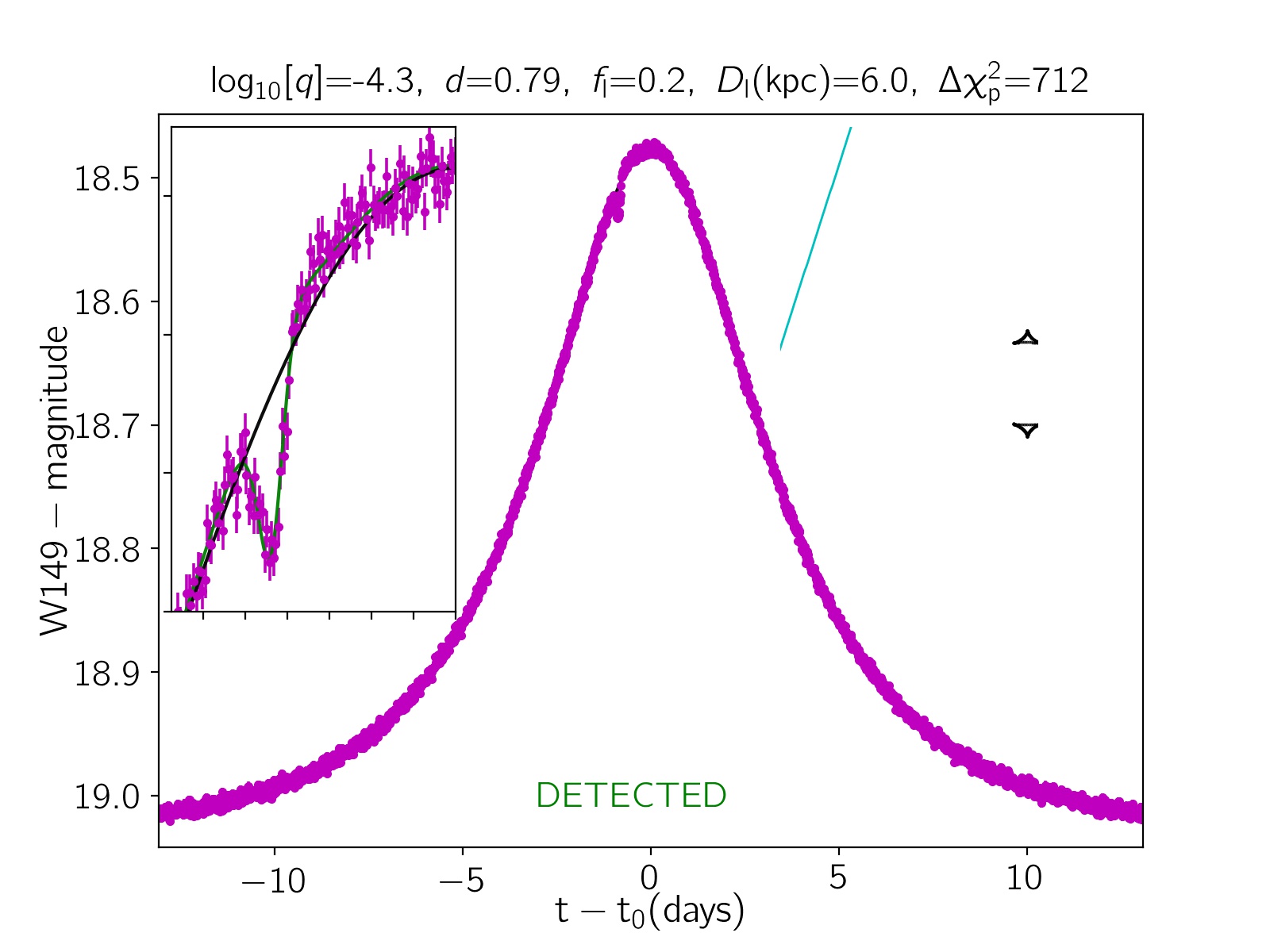}
\includegraphics[width=0.49\textwidth]{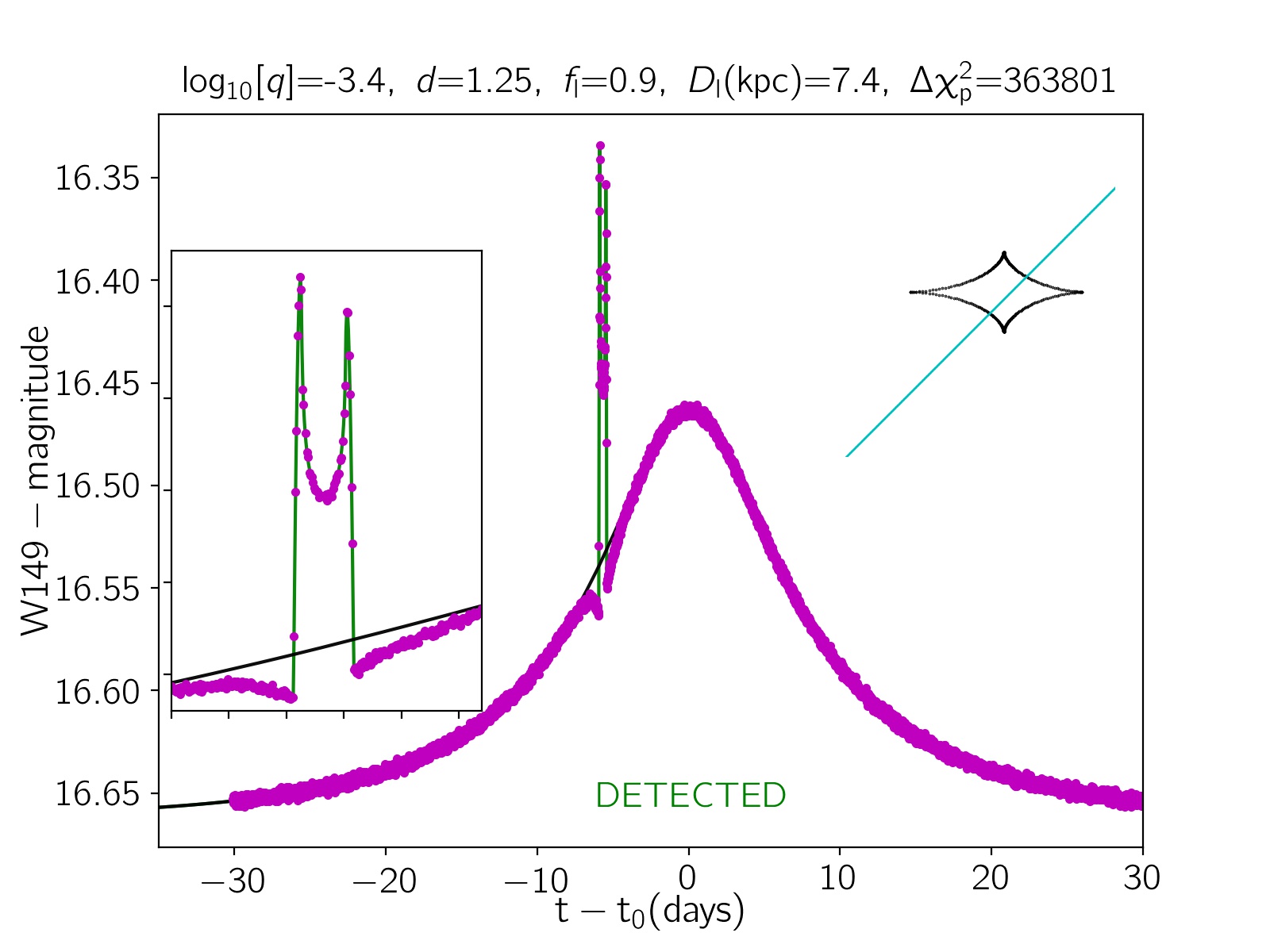}
\caption{Two simulated microlensing lightcurves (solid green curves) from habitable planetary systems detectable by the \wfirst~telescope. The solid black curves are the best-fitted single lightcurves. Magenta points display the simulated synthetic data points taken by the \wfirst~telescope. The source trajectories projected on the lens plane (cyan lines) and caustic curves (black curves) are shown in the right-hand insets. The parameters of these lightcurves are given at the top of the plots.}\label{lights}
\end{figure*}
\section{Detecting the HZ planets in the \wfirst~mission}\label{three}
The \wfirst~telescope will potentially detect many microlensing event toward the Galactic bulge during its mission \citep{Penny2019, Penney2020}. In order to accurately evaluate the \wfirst~ability for discovering planets in the HZs, a comprehensive Monte-Carlo simulation of such events is necessary, which will be done in this section. Many of the planetary microlensing lightcurves are created first. Then, we simulate synthetic observing data points for the generated lightcurves and finally extract the events with the detectable planetary signatures. These steps in the simulation are explained in the following. 

The simulation is done toward the subfields which will be observed by \wfirst.~The centers of these subfields, as shown in Fig. (7) of \citet{Penny2019}, are $(l^{\circ},~b^{\circ})=(1.3,~-0.89)$, $(0.9,~-0.89)$, $(1.3,~-1.63)$, $(0.9,~-1.63)$, $(0.5,~-1.63)$, $(0.1,~-1.63)$, and $(-0.3,~-1.63)$. The size of each sub-field is $\simeq 0.75\times 0.4 ~\rm{deg}^{2}$. Towards each given direction, the source distance, $D_{\rm s}$, is determined from the cumulative mass density versus distance from the observer in our galaxy:
\begin{eqnarray}
\frac{d^{2}M}{d\Omega~dD_{\rm s}} = D_{\rm s}^{2}~\sum_{i=1}^{4} \rho_{i}(D_{\rm s},~l,~b),
\end{eqnarray}
\noindent here, the summation is done over all structures, i.e., the Galactic bulge, thin and thick discs, and the stellar halo. The mass densities, $\rho_{i}$, are given by the Besan\c{c}on model \footnote{\url{https://model.obs-besancon.fr/}} \citep{Robin2003, Robin2012}. For each source star, we determine its absolute magnitude, mass, radius from different stellar populations of our galaxy as given by this model \citep[see, e.g.,][]{Sajadian2010Hot, 2015sajadianvv}. The source radius determines $\rho_{\ast}$ which is the projected source radius on the lens plane and normalized to the Einstein radius. $\rho_{\ast}$ significantly changes the magnification factor in high-magnification or caustic-crossing microlensing events \citep[][]{1994wittmoa, Nemiroff1994}. 

In the next step, the lens objects are simulated. The lens location is chosen from the range $[10\rm{pc},~D_{\rm s}]$, according to the microlensing event rate: 
\begin{eqnarray}
\Gamma(D_{\rm l}) \propto\sqrt{\frac{D_{\rm l}(D_{\rm s}-D_{\rm l} ) }{D_{\rm s}}}~\sum_{i=1}^{4} \rho_{i}(D_{\rm l},~l,~b),
\end{eqnarray}
where, the first factor is related to the Einstein radius. In this relationship and when characterizing the lens distance, we ignore variations in the lens-source relative velocity $v_{\rm{rel}}$ with respect to the lens distance. $v_{\rm{rel}}$ is the difference of the lens and source velocities as measured by the observer (in the observer coordinate system), while both velocities are projected onto the sky plane \citep[see, e.g., Eq. (2) of ][]{2021sajadians}. This relative velocity determines the Einstein crossing time, $t_{\rm E}=R_{\rm E}/v_{\rm{rel}}$, which specifies the timescale of microlensing events. Other parameters of the lens object, i.e., their photometric properties, mass, radius, effective temperature are taken from different stellar populations in the Besan\c{c}on model. We explain how the planets are characterized in subsections \ref{deteef} and \ref{static}.

The lens impact parameter is chosen uniformly from the range $u_{0} \in [0,~1]$. The angle between the source trajectory and the binary axis is taken from the range $\xi \in [0,~360]$ deg. Total observing time of the \wfirst~mission is $T_{\rm{obs}}=5$ yrs which includes $6$ $72$-day seasons. Three observing seasons will occur at the start of the mission with $110$-day seasonal gaps and three other seasons will occur at the end of the mission. We randomly take the time of the closest approach of simulated events, $t_{0}$, in the range of $[0,~T_{\rm{obs}}]$. To calculate the magnification factor in binary and planetary microlensing events we use the $\rm{RT}$-model \footnote{\url{http://www.fisica.unisa.it/gravitationastrophysics/RTModel.htm}} which was well developed by V.~Bozza \citep{Bozza2018, Bozza2010, Skowron2012}. 

\begin{table*}
\centering
\caption{The \wfirst~efficiencies for detecting planets within the HZs with different mass values and properties of these detectable events.}\label{tab1}
\begin{tabular}{ccccccccccccc}
\toprule[1.5pt]
$M_{\rm p}$&$\epsilon_{\rm{OHZ,~LS}}$&$\epsilon_{\rm{OHZ,~HS}}$&$\epsilon_{\rm{CHZ,~LS}}$&$\epsilon_{\rm{CHZ,~HS}}$&$\left<D_{\rm l}\right>$&$\left<M_{\rm h}\right>$&$\left<f_{\rm l}\right>$&$\left<m_{W149}\right>$&$\left<t_{\rm E}\right>$&$\left<d\right>$&$f_{\rm c}:f_{\rm i}:f_{\rm w}$ &  $f_{\rm{si}}$\\
&$[\%]$&$[\%]$&$[\%]$&$[\%]$&$\rm{(kpc)}$&$\rm{(M_{\odot})}$&&$\rm{(mag)}$&$\rm{(days)}$&& $[\%]$ & $[\%]$\\
\toprule[1.5pt]
$M_{\oplus}$ &$0.02$ & $0.01$ & $0.01$ & $0.01$ & $7.5$ & $0.83$ & $0.35$ & $18.0$ & $21.2$ & $0.78$ & $85$ : $1$ : $14$ & $5$\\
$10~M_{\oplus}$ &$0.14$ & $0.09$ & $0.13$ & $0.09$ & $7.2$ & $0.58$ & $0.26$ & $18.8$ & $18.3$ & $0.63$ & $84$ : $4$ : $12$ & $11$\\
$100~M_{\oplus}$ &$1.75$ & $1.5$ & $1.86$ & $1.60$ & $6.8$ & $0.31$ & $0.14$ & $19.2$ & $16.7$ & $0.28$ & $96$ : $2$ : $2$ & $4$\\
$M_{\rm J}$ &$5.20$ & $5.01$ & $5.47$ & $5.29$ & $6.7$ & $0.28$ & $0.12$ & $19.1$ & $17.7$ & $0.22$ & $97$ : $1$ : $2$ & $3$\\
$5~M_{\rm J}$ &$17.03$ & $16.98$ & $17.94$ & $17.89$ & $6.5$ & $0.28$ & $0.11$ & $19.0$ & $19.0$ & $0.17$ & $98$ : $1$ : $1$ & $2$\\
$10~M_{\rm J}$ &$25.43$ & $25.41$ & $26.79$ & $26.77$ & $6.5$ & $0.29$ & $0.10$ & $19.0$ & $19.7$ & $0.15$ & $98$ : $1$ : $1$ & $2$\\
\hline
\end{tabular}
\end{table*}

During a microlensing event, the angular Einstein radius, $\theta_{\rm E}$, is much smaller than the angular size of the FWHM (Full width at Half Maximum) of the target PSF (Point Spread Function), $\theta_{\rm PSF}$. In the \wfirst~microlensing survey, $\theta_{\rm PSF} \simeq 0.33$ arcs, i.e., around three times of the \wfirst~pixel size which is $0.11$ arcs \citep{Penny2019}. Hence, the lens flux is always blended with the source flux and we consider this effect in the simulation. In order to evaluate the blending effect owing to the lens flux, we define the lens blending factor as:
\begin{eqnarray}
f_{\rm l}=\frac{F_{\rm l}}{F_{\ast} +\sum_{i} F_{i}},
\end{eqnarray}
which is the ratio of the lens flux $F_{\rm l}$ to the total flux due to the source star itself $F_{\ast}$ and other blending stars (which their angular distances from the source star are smaller than the radius of the source PSF) at the baseline. The greater the $f_ {\rm l}$, the greater the contamination caused by the lens brightness.

Using the specified parameters, e.g., $\rho_{\ast},~t_{\rm E},~u_{0},  t_{0}, ...$, we make the microlensing lightcurves. For each simulated lightcurve, we generate synthetic data points taken during the \wfirst~observing program, as described in \citet{Bagheri2019, Sajadian2020}. The cadence of data is fixed at $15.16$ min. The photometric error bars of the synthetic data points, $\sigma_{\rm m}$, are specified based on the magnified apparent magnitudes of the targets, as shown in Figure (4) of \citet{Penny2019}. 

We first extract the detectable microlensing events from simulated events using three criteria which have been explained in \citet{Sajadian2019}. Then, we evaluate the detectability of the planetary signatures in extracted events. In order to realize the planetary signatures, we exert two extra criteria. Because the planet-host mass ratio, $q=M_{\rm p}/M_{\rm h}$, is small, the resulting planetary signatures from caustic-crossing features (or passing close to caustic curves) are small and short. For these events, the best-fitted single models are the real event by considering only primary lenses (ignoring their planets), which are located at $(-d~q/ (1+q),~0)$. Here, $d=s_{\mathcal{P}}/R_{\rm E}$ is the planet-host distance projected on the sky plane and normalized to the Einstein radius. Our criteria are as follows. (i) There are at least three consecutive data points that are $4 \sigma_{\rm m}$ above (or below) the best-fitted single models. (ii) $\Delta \chi^{2}_{\rm p}>300,~800$, corresponding to low and high sensitivities to planetary signatures (LS and HS), where $\Delta \chi^{2}_{\rm p}= \chi^{2}_{\rm{real}}-\chi^{2}_{\rm{single}}$, the difference between $\chi^{2}$ values obtained by fitting the real models and the best-fitted single-lens models. Here, we assume that the best-fitted planetary models are real models. We note that the threshold of $500$ for $\Delta \chi^{2}_{\rm p}$ is often adjudged as a robust criterion for detecting planetary signatures in microlensing events \citep[see, e.g., ][]{Gould2010ww, Udalski2018ww, 2018Ryu, 2020Hirao}. The first criterion ejects symmetric events that may be misinterpreted as single microlensing events, unless the deviated data points exactly occur at the time of the closest approach, which is rare. Our aim is to evaluate the detectability of planetary signatures and we do not pay attention to any other possible degenerate solutions.

We aim to respond two main questions from this simulation. (1) If there is an exoplanet with a given mass in the HZ, what is the \wfirst~efficiency for detecting this planet? Furthermore, for what lensing configurations is this efficiency maximized? (2) During the \wfirst~mission, how many habitable planets are detectable? On average what properties do these detectable planets within the HZs have? These two issues are referred to in the following subsections.

\subsection{Detection efficiency} \label{deteef}
Here, we study the \wfirst~efficiency for detecting HZ planets during its mission by doing a Monte-Carlo simulation. For each simulated microlensing event, we assume that one exoplanet is located in the OHZ of the lens object and choose the semi-major axis of its orbit uniformly from the range of $s\in [s_{\rm{OHZ},~i},~s_{\rm{OHZ},~o}]$. We assume the planet's orbit around its host star is circular and disregard the orbital motion effect of microlens systems \citep[see, e.g., ][]{OrbitalM}. Hence, for projecting the planetary orbit onto the sky plane, we need one inclination angle, $i$, as \citep[see, e.g.,][]{1992Gould, 2015Batista}:
\begin{eqnarray}
s_{\mathcal{P}} = s~\sin(i).\label{sps}
\end{eqnarray}
In fact, if we assume the host star at the center of a sphere, its orbiting planet can be at any random position over the surface of that sphere. Its location is specified with two angles in the spherical coordinate system which are the azimuthal and polar angles. The second one is the inclination angle of the planet orbit with respect to the line of sight toward the observer (polar axis). To uniformly choose initial position of the planet over the sphere surface, $\cos(i)$ is chosen uniformly from a range of $[0,~1]$ in the simulation. Therefore, on average the ratio of the projected planet-host distance to their real distance, $f_{\mathcal{P}}=\big<s_{\mathcal{P}}/ s\big>$, is 
\begin{eqnarray}
f_{\mathcal{P}}=\left<\sin (i)\right>=\int_{0}^{1} \sqrt{1-x^2}~dx=\frac{\pi}{4}, \label{projected}
\end{eqnarray}
where $x=\cos (i)$.

Two example simulated lightcurves are represented in Figure \ref{lights}. The planets are located in the HZs of their host stars. In these plots, the best-fitted and single-lens (solid black curves) and the real (solid green) lightcurves are displayed. The planetary signatures, as zoomed in the left-hand insets, are covered by the synthetic data points (magenta points) and are detectable. In the left and right panels, the planetary systems make close and wide caustic topologies, respectively. The source trajectories projected on the lens plane in these events (cyan lines) pass close to and over the planetary caustic curves. In the second lightcurve, although the blending effect due to the host lens is high, the planetary signature is detectable during its caustic-crossing features.  

We perform the Monte-Carlo simulation toward $(l,~b)=(1.0^{\circ},~-1.5^{\circ})$ for planet masses of $M_{\rm p}= M_{\oplus},~10~M_{\oplus},~100~M_{\oplus},~M_{\rm{J}},~5~M_{\rm J},~10~M_{\rm{J}}$, i.e., Earth-mass, Super-Earth, Saturn-mass, Jupiter-mass and Super-Jupiter planets (the index $\rm{J}$ refers to Jupiter). Accordingly, the inputs of the simulation are HZ planetary systems acting as microlenses and outputs are the ones that make the planetary microlensing events detectable by \wfirst.~In Table \ref{tab1}, we report the detection efficiency values, $\epsilon$, and the average parameters of output events for the different masses of planets as specified in different rows. The indexes LS and HS for the efficiencies refer to different criteria for detecting planetary signatures (i.e., the thresholds $300$ and $800$ for $\Delta \chi^{2}_{\rm p}$).  In this table, $m_{W149}=-2.5 \log_{10}\left[ F_{\ast}~+~F_{\rm l}~+~\sum_{i} F_{i} \right]$ is the apparent magnitude of the targets at the baseline. $f_{\rm c}$, $f_{\rm i}$, $f_{\rm w}$ and $f_{\rm{si}}$ are percentages of detectable planetary systems that make close, resonant, wide and close-to-resonant (with $d_{c}^{3}<d<d_{\rm w}^{1.8}$) caustic topologies, respectively.
\begin{figure*}
	\centering
	\includegraphics[width=0.49\textwidth]{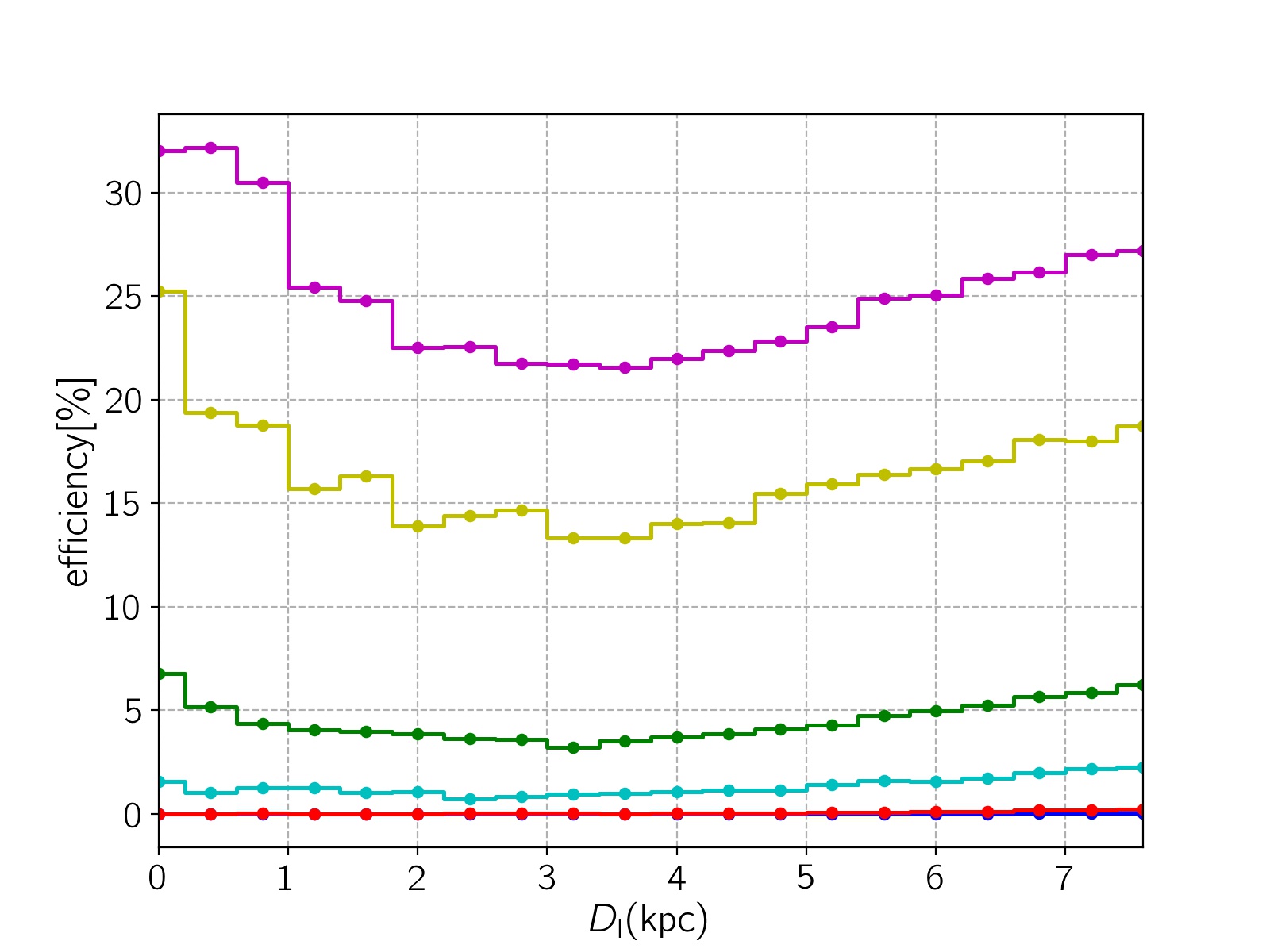}
	\includegraphics[width=0.49\textwidth]{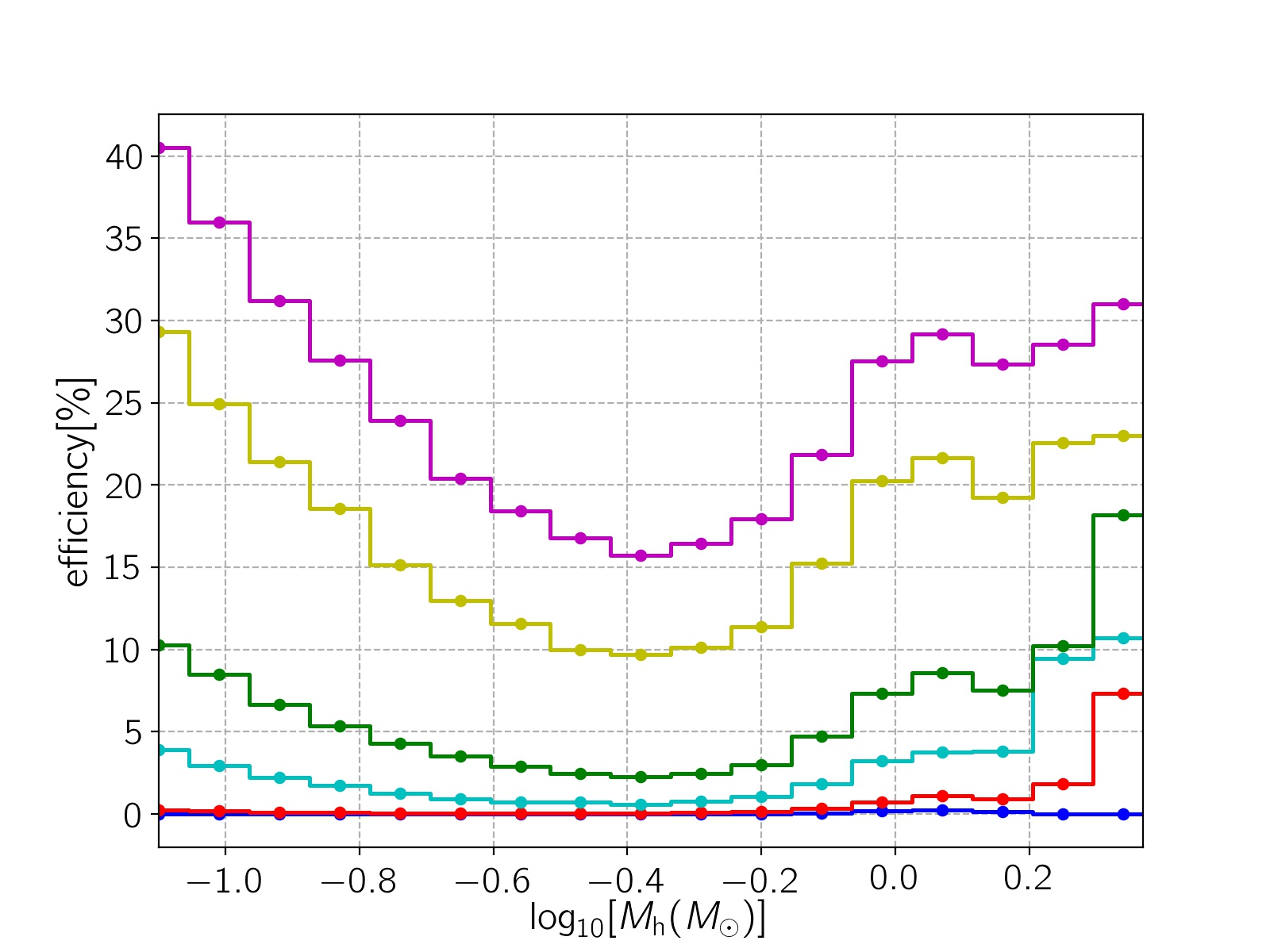}
	\includegraphics[width=0.49\textwidth]{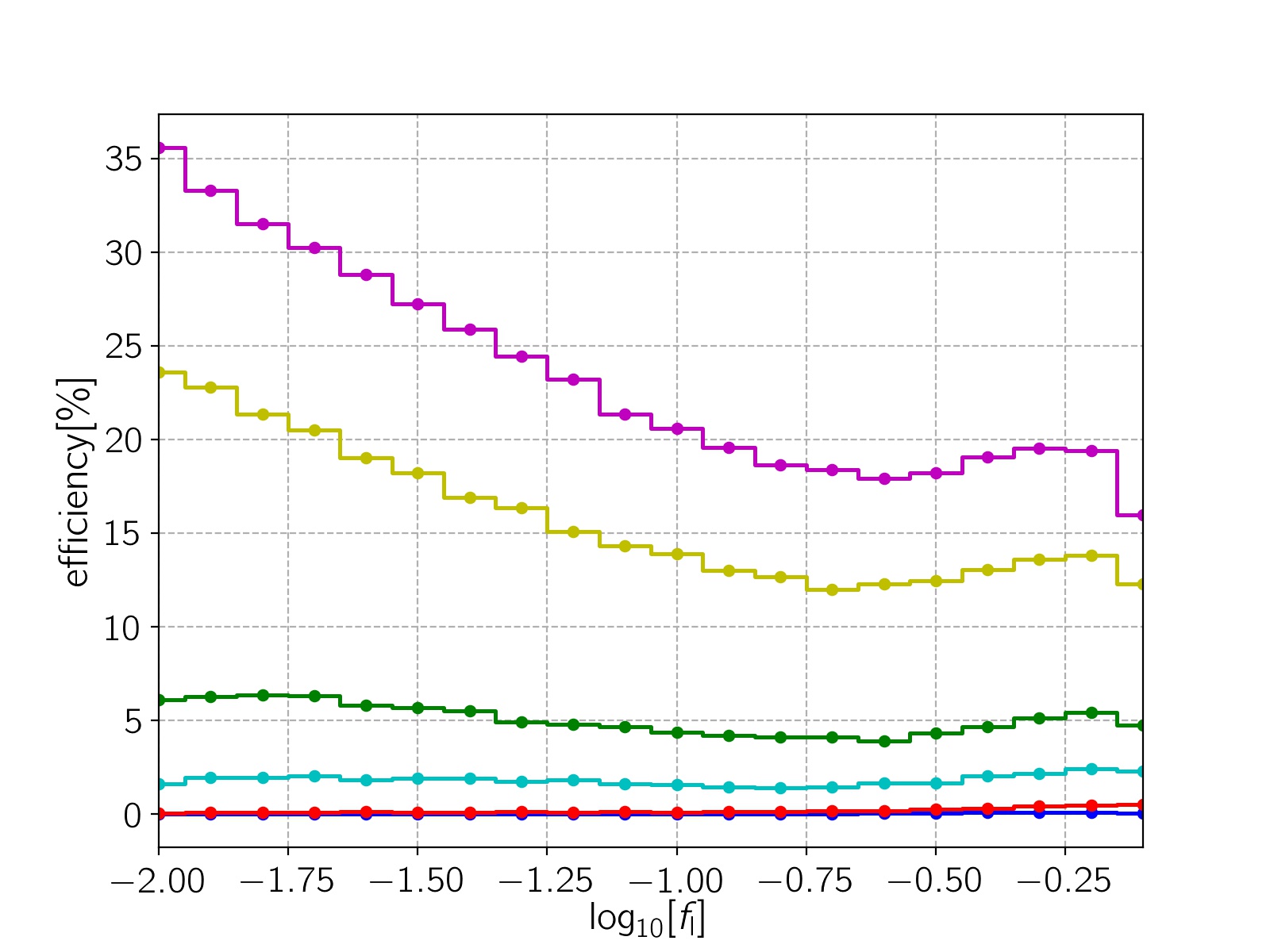}
	\includegraphics[width=0.49\textwidth]{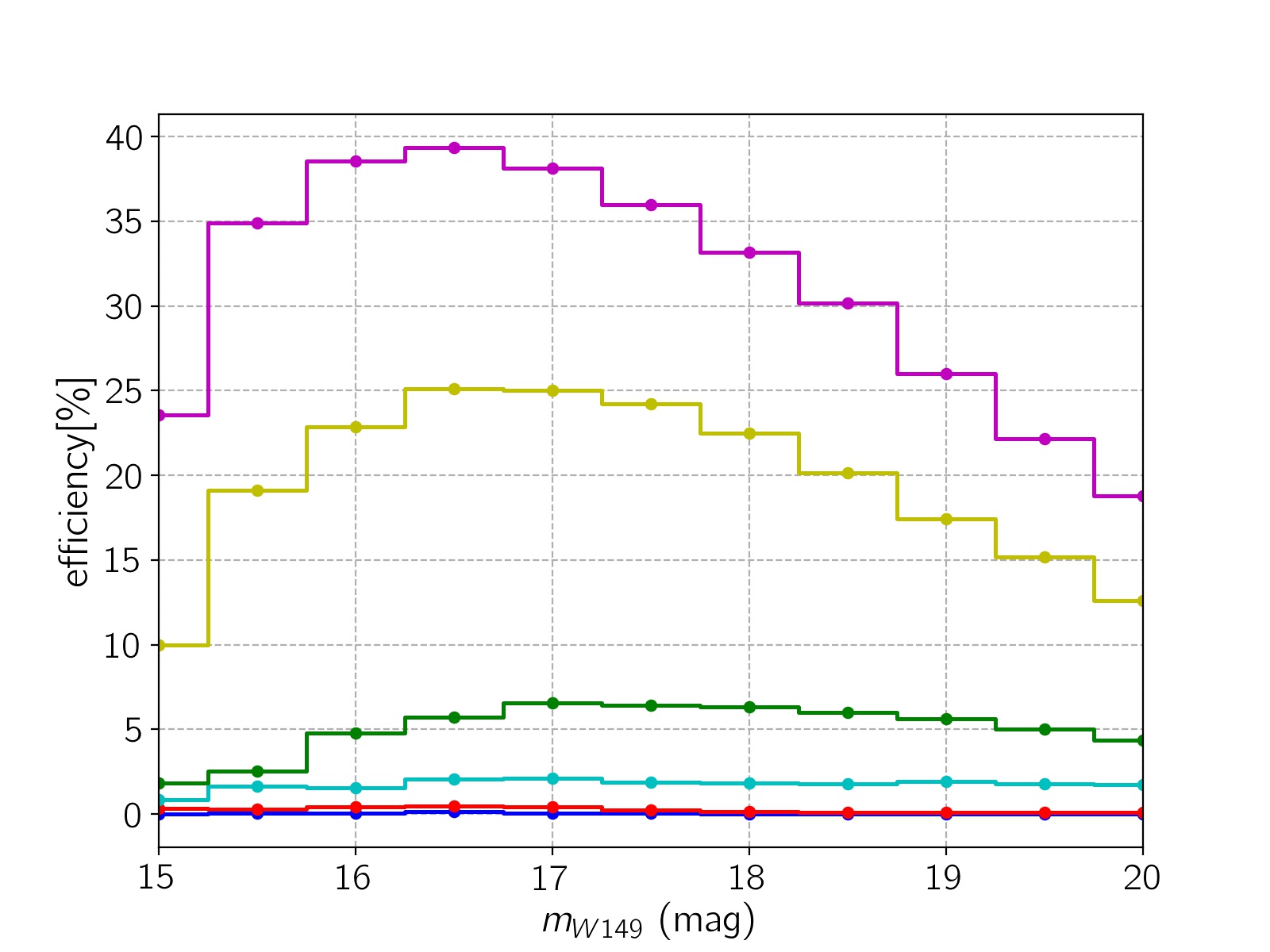}
	\caption{The \wfirst~efficiencies for detecting planets in the OHZs with LS criteria ($\epsilon_{\rm{OHZ},~\rm{LS}}$) in terms of the lens distances from the observer, the mass of the host lenses, the lens blending effect, and the target apparent magnitude at the baseline are shown in different panels. Each panel contains six efficiency curves which are corresponding to planets with masses $10~M_{\rm J}$ (magenta), $5~M_{\rm J}$ (yellow) $M_{\rm J}$ (green), $100~M_{\oplus}$ (cyan) and $10~M_{\oplus}$ (red) and $M_{\oplus}$ (blue).}\label{Effiplots}
\end{figure*}

\noindent Additionally, in Figure \ref{Effiplots} we show the dependence of the LS efficiencies for detecting planets in the OHZs, $\epsilon_{\rm{OHZ},~\rm{LS}}$, on the distance of the planetary systems from the observer $D_{\rm l}$, the mass of the primary lenses $M_{\rm h}$, the lens blending effect $f_{\rm l}$, and the target apparent magnitude at the baseline $m_{W149}$, represented in different panels. In each panel, six given curves are related to different planet masses, namely, $10~M_{\rm J}$ (magenta), $5~M_{\rm J}$ (yellow), $M_{\rm J}$ (green), $100~M_{\oplus}$ (cyan), $10~M_{\oplus}$ (red), and $M_{\oplus}$ (blue). We list some key points here.  
\begin{figure}
\centering
\includegraphics[width=0.49\textwidth]{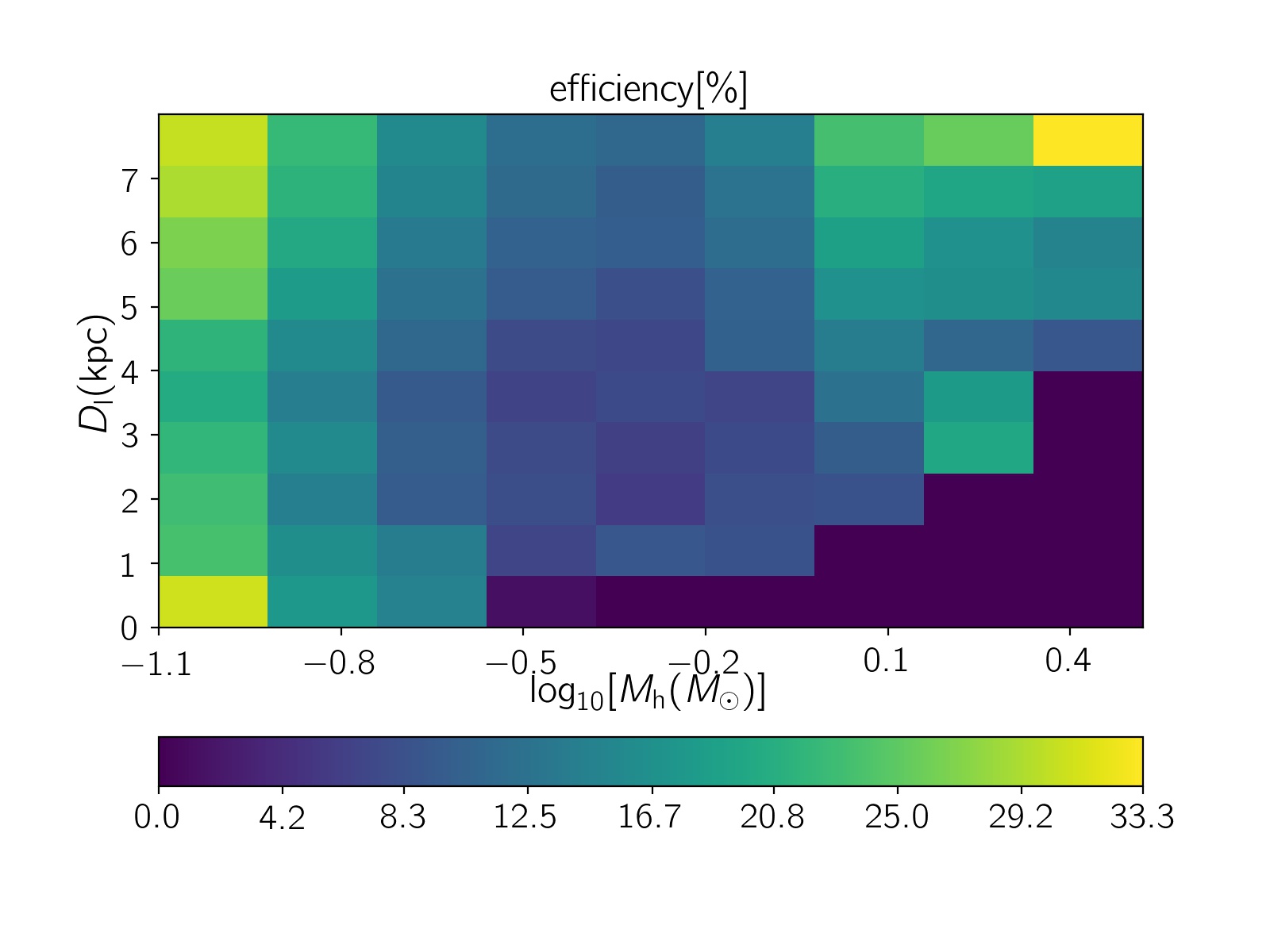}
\caption{Map of the LS \wfirst~efficiency for detecting Jupiter-mass planets around host stars with masses $M_{\rm h}$ at $D_{\rm l}$ from the observer.}\label{map}
\end{figure}

\begin{itemize}
\item Generally, the \wfirst~detection efficiency increases by enhancing the mass of planets. Their relation is almost linear for planets with the mass $M_{\rm p} \lesssim M_{\rm J}$. By increasing the planet mass by two orders of magnitude, from $M_{\oplus}$ to $100~M_{\oplus}$, the efficiency increases by two orders of magnitude ($0.01\%$ to  $1\%$) as well. But, it tends toward a constant value for the heavier planets. 
	
\item Only $\sim 0.01\%$ and $0.1\%$ of Earth and super Earth-mass planets in habitable zones are detectable through the \wfirst~microlensing observations. The properties of these detectable planetary events are: (i) Habitable Earth-mass planets are detectable when they rotate around G and K-type stars located at a distance beyond $7$ kpc from the observer (close to the Galactic bulge). (ii) The brightness of their primary lenses is considerable. (iii)  These planetary systems produce rather close or wide caustic topologies, owing to small values of $q$.

\item $\sim 5-25\%$ of Jupiter and super Jupiter-mass planets within habitable zones are detectable by \wfirst.~In these detectable events, the primary lens objects are most often M dwarfs (as usual), resulting less blending effect from the brightness of host objects. These planetary systems make mainly close caustic topologies. Jupiter-mass planets can have habitable moons. 

\item As mentioned in the previous section, the detection efficiency depends on $D_{\rm l}$ and it maximizes for the lens distances offering smaller Einstein radii. Generally, the \wfirst~efficiency for detecting HZ planets is maximized for three classes of planetary microlensing events with close caustic topologies, which are pointed out here.

\noindent (a) The events with massive primary lenses, $M_{\rm h}\gtrsim 0.6 M_{\odot}$ while their mirolens systems are very far, $D_{\rm l}\gtrsim 7$ kpc. In these events and for Jupiter-mass planets within the HZs, the mass ratio is very small $q \lesssim 0.001$, but $d \gtrsim 0.17$. The parallax amplitude is very small, $\pi_{\rm{rel}}\lesssim 0.02$ mas. Here, $\pi_{\rm{rel}}=\rm{AU}(1/D_{\rm l}-1/D_{\rm s})$. These planetary systems have orbital periods as long as $P \gtrsim 100$ days. Therefore, the orbital motion effect due to lenses is likely small.
	
\noindent The events with low-mass primary lenses, $M_{\rm h} \lesssim 0.1 M_{\odot}$ while their microlens systems are either (b) very close to the observer $D_{\rm l}\lesssim 1$ kpc or (c) very close to the Galactic bulge $D_{\rm l}\gtrsim 7$ kpc. By assuming a Jupiter-mass planet in the HZ of the primary lenses, these events (in both classes) have $q \gtrsim 0.01$ and $d \lesssim 0.04$. For class (b) of microlensing events, the parallax effect is significant with $\pi_{\rm{rel}}\gtrsim 0.87$ mas. These planetary systems have the short orbital periods with $P\lesssim 7$ days, which can potentially make considerable orbital motion effects on caustic-crossing features.

\noindent For that reason, we have two peaks in the two first panels of Figures \ref{Effiplots}. In Figure \ref{map} we depict the map of the \wfirst~efficiency for detecting Jupiter-mass planets around primary lenses (to better display this point). Three local maxima in this parameter space $M_{\rm h}$-$D_{\rm l}$ are manifest. 

\noindent We note that the events in the classes (a) and (c) occur more frequently. The horizontal sizes of central and planetary caustics (over the binary axis) in close caustic topologies are proportional to $\Delta_{\rm c} \propto 4q/(d-1/d)^{2}$ and $\Delta_{\rm p} \propto \sqrt{q}~d^{3}$ \citep{2000Bozza, 2005An, Chung2005}. Hence, central and planetary caustics of the class (a) events are larger (by one and two orders of magnitude, respectively).  

\item The maximum detection efficiency occurs when the targets are as bright as $m_{W149}\sim 16$-$17$ mag (at the baseline). A significant part of this apparent brightness of the targets at the baseline is owing to the brightness of the primary lenses. 
\end{itemize}

\begin{table*}
\centering
\caption{The properties and statistics of planetary microlensing events within the OHZs and CHZs which are detected by \wfirst~ with the LS and HS criteria.}\label{tab2}
\begin{tabular}{cccccccccccccccc}\toprule[1.5pt]
$~$&$\left<M_{\rm p}\right>$&$\left<D_{\rm l}\right>$&$\left<M_{\rm h}\right>$&$\left<f_{\rm l}\right>$&$\left<m_{W149}\right>$&$\left<t_{\rm E}\right>$&$\left<d\right>$&$\left<R_{\rm E}\right>$&$f_{\rm c}:f_{\rm i}:f_{\rm w}$&$f_{\rm{si}}$ &$\epsilon_{\rm p}$&$\epsilon_{\rm{HZ}}$&$N_{\rm HZ}$&$\epsilon_{\rm{HZ},~\mathcal{P}}$&$N_{\rm HZ,~\mathcal{P}}$\\
&$M_{\rm J}$&$\rm{(kpc)}$&$\rm{(M_{\odot})}$&&$\rm{(mag)}$&$\rm{(days)}$&&$\rm{(AU)}$&$[\%]$& $[\%]$&$[\%]$&$[\%]$&&$[\%]$&\\ 
\toprule[1.5pt]	
\multicolumn{16}{c}{$\rm{OHZ}~\rm{planetay}~\rm{Microlensing}$}\\
$\rm{LS}$&$2.71$&$6.89$&$0.51$&$0.24$&$19.5$&$23.1$&$0.29$&$1.63$&   $96$ : $3$ : $1$ & $5.1$ &$3.00$&$3.35$&$27$&$4.29$&$35$\\ 
$\rm{HS}$&$2.76$&$6.90$&$0.53$&$0.25$&$19.3$&$24.7$&$0.30$ &$1.70$& $95$ : $4$ : $1$& $6.0$ &$2.79$ &$3.13$&$24$&$4.07$&$31$ \\ 
\hline 
\multicolumn{16}{ c }{$\rm{CHZ}~\rm{planetay}~\rm{Microlensing}$}\\
$\rm{LS}$ &$2.57$ &$6.63$ & $0.43$ & $0.22$ & $19.5$ & $18.4$ & $0.31$ & $1.43$  & $95$ : $3$ : $2$ & $7.2$ & $3.00$ & $1.79$ &$15$ & $2.76$ &$22$\\
$\rm{HS}$&$2.60$ & $6.69$& $0.44$ & $0.23$ & $19.4$ & $19.2$ & $0.33$ & $1.47$  & $94$ : $4$ : $2$ & $8.4$ & $2.79$ & $1.65$ &$12$ & $2.62$ &$20$ \\
\hline
\end{tabular}
\end{table*}

\subsection{Statistical Calculations}\label{static}
In order to estimate the number of habitable planets that could potentially be detected during the \wfirst~mission, we will do a more realistic simulation of all possible planetary systems detectable by this telescope and extract the events with planets in the HZs.  

For doing this realistic simulation, we choose semi-major axis of the planetary orbit, $s$, from the \"{O}pik's law \citep{Opik1924}, i.e., 
\begin{eqnarray}
dN/ds \propto 1/s,
\end{eqnarray} 
and from the range $s \in [0.01,~100]$ AU. To extract the lens mass ratio, we use the mass ratio function which has been resulted from microlensing observations. One of the advantages of microlensing observations is studying the abundance of bound planets in planetary systems. For instance, \citet{Cassan2012} have found that $17\%$ of stars have Jupiter-mass planets and low-mass planets are more common. From the MOA microlensing observations, \citet{Suzuki2016} have discovered a turnover in the mass ratio function at $q\simeq 1.7\times 10^{-4}$. Their results have been confirmed by \citet{Udalski2018ww} for a larger sample and they have offered a twofold mass ratio function which is given by:  
\begin{eqnarray}
\frac{dN}{d\log_{10} q} \propto~
\begin{cases} 
q^{0.73} & q \lesssim 2\times 10^{-4},\\
q^{-0.93} &  q > 2 \times 10^{-4}. \\
\end{cases} 
\end{eqnarray}
In the simulation, we choose the mass ratio using this function and from the range $q \in [10^{-5},~0.06]$. However, we disregard the events with  $M_{\rm p}(=q M_{\rm h})>15 M_{\rm J}$ which are brown dwarfs. In the simulation, we extract planetary microlensing events which (i) have detectable planetary signatures by applying LS and HS criteria, and (ii) their planets are within the HZs. We perform the simulation toward all mentioned sub-fields and the average parameters of these events are mentioned in Table \ref{tab2}. 

\noindent We conclude that the fraction of simulated microlensing events with detectable planetary signatures (by exerting LS criteria) during the \wfirst~mission is $\epsilon_{\rm p}=3.0\%$. The fractions of these planetary events that their planets locate in the OHZs and CHZs (i.e., $s \in [s_{\rm{HZ},~i},~s_{\rm{HZ},~o}]$) are $\epsilon_{\rm{HZ}}\simeq 3.35,~1.79\%$, respectively.

In order to estimate the number of habitable planets that can be detected by \wfirst,~we use predictions of detectable microlensing events by this telescope, done by \citet{Penny2019}. Accordingly, \wfirst~will detect $N_{\rm e}\sim 27000$ microlensing events. If we assume that each of these microlenses has one planet orbiting it at a distance of $[0.01,~100]$ AU (specified by the \"{O}pik's law), the number of habitable planets which are detectable by \wfirst~is evaluated as $N_{\rm HZ} \simeq N_{\rm e}~\epsilon_{\rm p}~\epsilon_{\rm{HZ}}$. Hence, by considering LS criteria \wfirst~will detect $27$ exoplanets in the OHZs, where $15$ of them are in the CHZs. If there is more than one exoplanet for each microlens, the total number of habitable exoplanets will increase. In fact, the simulations in this work are based on the Besan\c{c}on model, similar to ones done by \citet{Penny2019}. We note that the \wfirst~efficiencies (given by our simulations) are the average values due to all \wfirst~observing directions. If we had had $N_{\rm e}(l,~b)$ for each line of sight, by calculating the efficiencies in different line of sights we could have determined $N_{\rm HZ}$ more accurately. However, $N_{\rm HZ}$ given in this paper is an acceptable estimation, because the \wfirst~observing directions are close to each other and the variations of related parameters are small. For instance, the average extinction in W149 towards the mentioned sub-fields are $\left<A_{\rm{W149}}\right>=1.58,~1.97,~0.79,~0.69,~0.79,~0.86,~0.94$, and the average optical depth for the observations toward these sub-fields are $\left<\tau(\times 10^{6})\right>=10.4,~10.5,~8.5,~9.7,~10.6,~10.2,~9.9$.

\noindent By applying HS criteria for detectability of planetary signatures, the number of detectable planets in the OHZs and CHZs is $24$ and $12$, respectively.  

These planets are most often rotating around $\rm{K}$-type stars which are on average at the distance farther than $6.7$ kpc from us. Targets in these events are on average as bright as $19.5$ mag in the W149 filter at the baseline. Some part of this brightness is due to the primary lens. Because primary lenses have a high brightness (around $0.24$ of the source brightness), follow-up observations can resolve primary lens objects. Detecting lens objects directly helps to resolve microlensing degeneracy and inferring parameters of planets. 

\noindent Owing to the brightness of primary lens objects, around $10.5\%$ of source stars in our sample (of detectable planetary events by \wfirst)~are giants. In current microlensing observation toward the Galactic bulge, this fraction is higher, $\sim 18.4\%$ \citep[see, e. g.,][]{2015Wyrzykowski, 2015sajadian}. 

From our simulation, we found that $20.7\%$ of detectable planetary microlensing events from habitable planetary systems have obvious caustic-crossing features (the events with the minimum source distance from caustic lines less than $5 \rho_{\ast}$, where $\rho_{\ast}$ is the projected source radius normalized to the Einstein radius). In other events, source stars pass close to the caustic instead of crossing it, e.g., the left event in Figure \ref{lights}. Some examples of such events are reported in \citet{Hanetal.2021, 2021Kondo}. 

\section{Discussion}\label{four}
In planetary microlensing events, one can measure the planet-host mass ratio and planet-host distance projected on the sky plane. The mass of primary components $M_{\rm h}$ is carefully evaluated in caustic-crossing microlensing events, if parallax and finite-size effects are measured \citep[see, e.g.,][]{2004Yoo, 2019Shvartzvald, 2018ApJHan}. But, microlensing observations give the projected planet-host distance onto the sky plane, $s_{\mathcal{P}}$, which is given by Equation \ref{sps}. The projected planet-host distance is always less than the actual distance, i.e., $s_{\mathcal{P}}<s$. 

\begin{figure}
\centering
\includegraphics[width=0.49\textwidth]{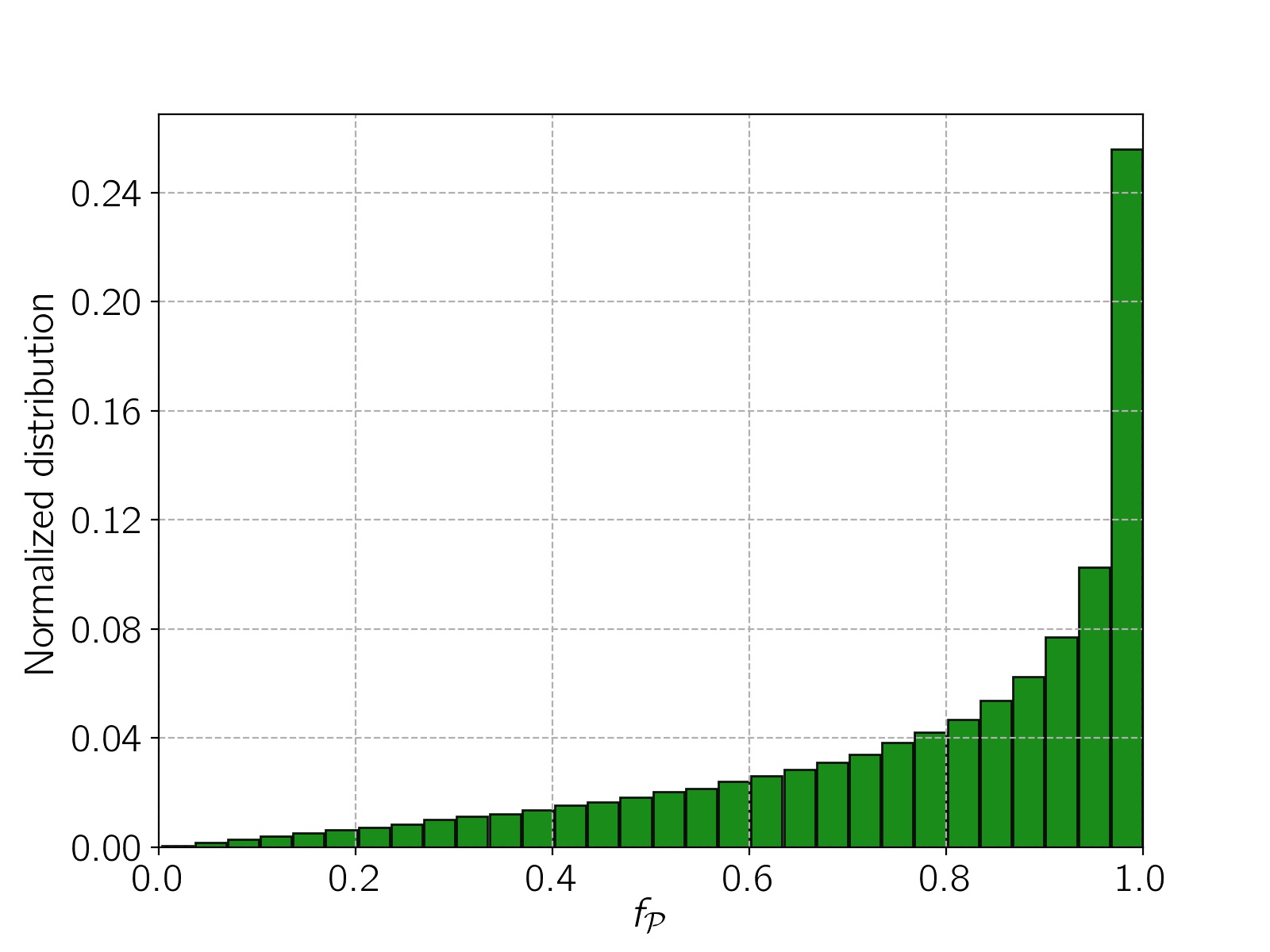}
\caption{The normalized distribution of $f_{\mathcal{P}}=s_{\mathcal{P}}/s$ for different values of the projection angle.}\label{distwe}
\end{figure}

\noindent For a given $s$, we choose $\cos(i)$ randomly. The distribution of resulting projected distances normalized to real distance, $s_{\mathcal{P}}/ s$, is shown in Figure \ref{distwe}. On average, $\left<s_{\mathcal{P}} \right>=0.79~s$, as given by Equation \ref{projected}. Hence, if microlensing observations find a planetary system with the projected distance in the habitable zone, the real distance between the planet and its host star is most likely larger than the measured amount. Since microlensing is more sensitive to normalized planet-host distances larger than those owing to habitable planetary systems, so microlensing observations overestimate the number of HZ planets. In the simulation of section \ref{three}, we found that fraction of detectable planetary microlensing events (by considering LS criteria) for which $s_{\mathcal{P}}$ is located in the OHZs and CHZs are $\epsilon_{\rm{HZ},~\mathcal{P}}=4.3,~2.8\%$, respectively. These efficiencies are greater than $\epsilon_{\rm{HZ}}$. Accordingly, \wfirst~will detect $N_{\rm{HZ},~\mathcal{P}}=35$ exoplanets with projected distances in the OHZs, which $22$ of them have projected distances in CHZs, by applying LS criteria. These numbers are given in the two last columns of Table \ref{tab2} as well.

If we consider only Earth-mass and super-Earth planets, fractions of total detectable planetary microlensing events by \wfirst~that (i) the planet mass is $M_{\rm p}<10 M_{\oplus}$ (Earth and Super Earth-mass planets) and (ii) their projected positions are within the OHZs and CHZs are $\epsilon_{\rm{HZ},~\mathcal{P},~\oplus}\simeq 0.11,~0.05\%$, respectively. Hence, only  $\sim 1$ of total detectable exoplanets in the OHZs (i.e., $35$) has $M_{\rm p} < 10 M_{\oplus}$. 

The orbital motion effect should be very small in detectable planetary microlensing events with planets in the HZs. There are two reasons. (i) In these events, the caustic-crossing features are rare and in most cases, caustic sizes are very small and source stars are passing close to caustic curves.  (ii) Most detectable events are from the second class of microlensing events introduced in subsection \ref{deteef}, i.e., the events with $D_{\rm l} \gtrsim 7$ kpc and $M_{\rm h} \gtrsim 0.6 M_{\odot}$ with the orbital period $P \gtrsim 100$ days. These long orbital periods likely have a small effect on microlensing lightcurves.

\section{conclusions}\label{five}
Detecting planets in the HZs is an important stage for probing extraterrestrial life, which is one of the active research areas in astrophysics and astronomy. One method for detecting exoplanets is microlensing observations. Although the sensitivity of microlensing observations to habitable planets is low, by this method one can probe planetary systems farther from the Earth position than those detectable through transit or radial velocity measurements. Hence, microlensing is a complementary method for detecting habitable planets in terms of their distances from us.   

In this work, we evaluated the \wfirst~efficiency to detect planets within the HZs by doing a Monte-Carlo simulation. We concluded that the LS \wfirst~efficiencies for detecting planets in the HZs were $0.01,~5,~25\%$ for Earth, Jupiter, and Super Jupiter-mass planets, respectively. This efficiency increases linearly with the mass, for $M_{\rm p} \lesssim M_{\rm J}$. But, it tends toward a constant value for the heavier planets. Earth-mass planets within the HZs are detectable around massive host stars with $\left<M_{\rm h}\right>\simeq 0.8 M_{\odot}$ at the distances farther than $7$ kpc from us. 

The \wfirst~efficiency for detecting planets in the HZs is maximized for three classes of planetary microlensing events with close caustic topologies. 

\noindent (a) The events with $D_{\rm l} \gtrsim 7$ kpc, $M_{\rm h}\gtrsim 0.6 M_{\odot}$ (massive primary lenses). By assuming Jupiter-mass planets located within the HZs of these lenses, these events have $q\lesssim 0.001$ and $d\gtrsim 0.17$ with negligible effects due to parallax and orbital motion of lenses (Their orbital periods are longer than $100$ days).

\noindent The events with low-mass primary lenses, $M_{\rm h} \lesssim 0.1 M_{\odot}$, while their lens systems are either (b) close to the observer with $D_{\rm l}\lesssim 1$ kpc (which have considerable parallax effects with the amplitude $\pi_{\rm{rel}}>0.87$ mas) or (c) close to the Galactic bulge, $D_{\rm l}\gtrsim 7$ kpc. For Jupiter-mass planets located in the HZs of primary lenses, the events in these two classes have $q\gtrsim 0.01$ and $d\lesssim 0.04$.  

\noindent In all of these microlensing events (three classes), planetary systems make close caustic topologies. The events in class (a) occur more frequently and make planetary and central caustics with larger sizes.  

We simulated an ensemble of detectable planetary microlensing events by \wfirst~and concluded that in $4\%$ of these events, the projected planet-host distances are in the OHZs. We noted that in $3\%$ of them, their planets are in reality in the OHZs. \wfirst~during its $5$-year mission will potentially detect $35$ planetary microlensing events with the projected planet-host distances, $s_{\mathcal{P}}$, in the OHZs. One of them will have $M_{\rm p}<10 M_{\oplus}$. We note that any difference between the predicted and observed number of planets in the HZ would also provide precious constraints either for planet formation theories and their abundance, or for the understanding of Roman photometry, or understanding the Galactic models.

\section*{Acknowledgements}
I thank the Department of Physics, Chungbuk National University, and especially C.~Han for hospitality. I am grateful to A.~Gould for his helpful discussion and the referee for his/her useful comments and suggestions.

\section*{DATA AVAILABILITY}
The data underlying this article will be shared on reasonable request to the corresponding author.

\bibliographystyle{mnras}
\bibliography{paperref}
\end{document}